\documentclass[twocolumn]{aastex63}
\usepackage[utf8]{inputenc}
\usepackage{placeins}
\usepackage{graphicx}
\usepackage{amsmath}
\usepackage{physics}
\usepackage{tikz}
\usetikzlibrary{calc,intersections,through,backgrounds,decorations.pathreplacing,arrows.meta,angles}

\usepackage{dcolumn}
\newcolumntype{d}[1]{D{.}{.}{#1}}
\newcommand*\lateraleye{%
       \scalebox{0.15}{
    \tikzset{every picture/.style={line width=0.75pt}} 
    \begin{tikzpicture}[x=0.75pt,y=0.75pt,yscale=-1,xscale=1]
    \draw  [line width=1.5]  (300,100.33) .. controls (326,122) and (352,135) .. (378,139.33) .. controls (352,143.67) and (326,156.67) .. (300,178.33) ;
    \draw  [fill={rgb, 255:red, 0; green, 0; blue, 0 }  ,fill opacity=1 ] (308.94,116.33) .. controls (313.87,116.33) and (317.86,125.51) .. (317.85,136.83) .. controls (317.84,148.15) and (313.84,157.33) .. (308.91,157.33) .. controls (303.99,157.32) and (300,148.14) .. (300.01,136.82) .. controls (300.02,125.5) and (304.02,116.32) .. (308.94,116.33) -- cycle ;
    \draw  [draw opacity=0][line width=1.5]  (314.84,166.6) .. controls (311.87,164.64) and (309.14,162.18) .. (306.76,159.24) .. controls (295.12,144.82) and (296.6,124.33) .. (310.07,113.45) .. controls (311.48,112.32) and (312.96,111.33) .. (314.5,110.49) -- (331.14,139.55) -- cycle ; \draw  [line width=1.5]  (314.84,166.6) .. controls (311.87,164.64) and (309.14,162.18) .. (306.76,159.24) .. controls (295.12,144.82) and (296.6,124.33) .. (310.07,113.45) .. controls (311.48,112.32) and (312.96,111.33) .. (314.5,110.49) ;
    \draw  [fill={rgb, 255:red, 255; green, 255; blue, 255 }  ,fill opacity=1 ] (304.43,124.2) .. controls (306.09,124.25) and (307.32,128.01) .. (307.18,132.6) .. controls (307.05,137.19) and (305.59,140.88) .. (303.93,140.83) .. controls (302.27,140.78) and (301.03,137.02) .. (301.17,132.43) .. controls (301.31,127.83) and (302.76,124.15) .. (304.43,124.2) -- cycle ;
    \end{tikzpicture}
    }\,}
\def\MarkRightAngle[size=#1](#2,#3,#4){
  \draw[black!80!black] ($(#3)!#1!(#2)$) -- ($($(#3)!#1!(#2)$)!#1!90:(#2)$) -- ($(#3)!#1!(#4)$)
}
\newcommand{\tikzAngleOfLine}{\tikz@AngleOfLine}                               
  \def\tikz@AngleOfLine(#1)(#2)#3{%                                            
  \pgfmathanglebetweenpoints{%                                                 
    \pgfpointanchor{#1}{center}}{%                                             
    \pgfpointanchor{#2}{center}}                                               
  \pgfmathsetmacro{#3}{\pgfmathresult}%                                        
  }

\begin{document}

\title{Two Brief Geometric Derivations of the Shklovskii Effect in Pulsars}

%add orcid ids
\author[0000-0002-7746-8993]{Thomas Donlon II}
\affiliation{Department of Physics and Astronomy, University of Alabama in Huntsville, 301 North Sparkman Drive, Huntsville, AL 35816, USA}
\correspondingauthor{Thomas Donlon II}
\email{thomas.donlon@uah.edu}

\begin{abstract}
    I provide two short derivations for the Shklovskii effect, which describes the apparent secular period drift of a periodic source due to its motion on the sky. These derivations use easily visualizable geometry and calculus, and might be useful for those who wish to gain intuition for the subject. \\\vspace{0.5cm}
\end{abstract}

\section{Background}

The Shklovskii effect describes the apparent secular period drift (or equivalently, acceleration) of a pulsar due to its proper motion on the sky. (Note that this effect is present for \textit{any} periodic astronomical source, although it is primarily discussed in the context of pulsars since it can be comparable to the observed period drift). It is written equivalently in terms of the tangential velocity ($v_\perp$) or proper motion ($\mu$) as \begin{equation} \label{eq:def}
    \frac{\dot{P}}{P} = \frac{v_\perp^2}{c d} = \frac{\mu^2 d}{c}.
\end{equation}

This effect was originally provided in \cite{Shklovskii1970} as a way to explain the observed secular period drifts in pulsars, who claimed that it was ``easily shown'' and did not provide a proof. Subsequent publications appear to either take this statement as fact, or derive it from a fairly involved argument using the derivative of the unit radial vector to a pulsar (see Section 2 and Appendix A of \citealt{Liu2018} for an example of this, which I will not describe in detail here). 

However, this way of deriving the result is fairly opaque, and it is not how I personally think about the problem. I am sharing here two derivations of this effect that only use geometry and (basic) calculus to arrive at the usual expression, in hopes that it will be helpful for students or those who wish to gain intuition for this formula.

\section{Helpful Machinery}

The Doppler effect provides a relationship between the frequency of an observed periodic signal ($f$) and the velocity of the emitting object along the observer's line of sight ($v_{||}$): \begin{equation}
    \Delta f = \frac{-\Delta v_{||}}{c}f.
\end{equation} (Note the flipped sign from the usual convention due to our choice of direction for $v$). By dividing both sides of this expression by a small interval of time $\Delta t$, we can convert this into an acceleration: \begin{align}
    -\frac{1}{f}\frac{\Delta f}{\Delta t} &= \frac{1}{c}\frac{\Delta v_{||}}{\Delta t}, \\ \nonumber
    -\frac{\dot{f}}{f} = \frac{\dot{P}}{P} &= \frac{a}{c}.
\end{align} Therefore, from Equation \ref{eq:def} it is satisfactory to show that \begin{equation}
    a = \frac{v_\perp^2}{d}
\end{equation} to derive the Shklovskii effect.

\section{Derivation One}

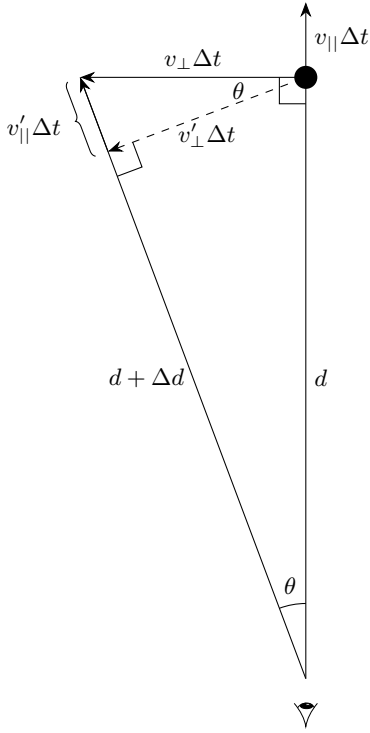
\begin{figure}
\centering
\begin{tikzpicture}
\coordinate (A) at (0,0);
\coordinate (B) at (0,8);
\draw [name path=A--B] (A) -- node[right] {$d$} (B);
\coordinate (C) at (-3,8);
\node[fill=black,circle] at (B) {};
\draw [name path=A--C] (A) -- node[left] {$d+\Delta d$} (C);
\draw [-{Stealth[length=2mm]},name path=B--C] (B) -- node[above] {$v_{\perp}\Delta t$} (C);
\draw [{Stealth[length=2mm]}-,dashed,name path=intersection] ($(A)!(B)!(C)$) -- node[below] {$v'_{\perp}\Delta t$} (B);
\MarkRightAngle[size=10pt](C,B,A);
\MarkRightAngle[size=10pt](A,$(A)!(B)!(C)$,B);
\draw [decoration={brace,raise=5pt},decorate] ($(A)!(B)!(C)$) -- node[left=0.8cm,below=-0.1cm] {$v'_{||}\Delta t$} (C);
\draw [-{Stealth[length=2mm]}] ($(A)!(B)!(C)$) -- (C);

\coordinate (D) at (0,9);
\draw [-{Stealth[length=2mm]},name path=B--D] (B) -- node[right] {$v_{||}\Delta t$} (D);

\tikzAngleOfLine(A)(B){\AngleStart}                                          
\tikzAngleOfLine(A)(C){\AngleEnd}                                            
\draw (A)+(\AngleStart:1cm) arc (\AngleStart:\AngleEnd:1cm) node[above=0.3cm,right=-0.05cm] {$\theta$};  
%\tikzMarkAngle{(A)}{(C)}{(B)}

\node [below=0.175cm,left=0.7cm] at (B) {$\theta$};

\filldraw[black] (0,-0.5) circle (0pt) node[rotate=-90]{$\lateraleye$};
\end{tikzpicture}
\caption{Schematic for the system. Not to scale. Arrows corresponding to velocities have been multiplied by the time interval $\Delta t$ to give them units of distance.} \label{fig:schematic}
\end{figure}

Let $d$ be the distance to the pulsar at time $t=t_0$. At time $t=t_0+\Delta t$, where $\Delta t$ is some short interval of time, the pulsar will be located a distance $d+\Delta d$ away from the observer. The pulsar's initial line-of-sight velocity contributes a fixed Doppler shift of the observed frequency, and can be ignored.

We use an unprimed coordinate system to represent the velocities of the pulsar at the initial time and a primed coordinate system to represent the velocities of the pulsar at time $t=t_0 + \Delta t$. 

Note that the initial tangential velocity ($v_\perp$) is perpendicular to the line-of-sight at time $t=t_0$, but is slightly offset (by an angle $\theta$) from the tangential velocity ($v'_\perp$) at the time $t=t_0+\Delta t$. Due to this offset, within the small time interval $\Delta t$ the pulsar has a slight perceived line-of-sight motion as well as a perceived motion on the sky. We can calculate the perceived acceleration as the change in the perceived line-of-sight motion, namely \begin{equation}
    \frac{\Delta v_{||}}{\Delta t} = a.
\end{equation} Note that \begin{equation}
    \Delta v_{||} = v'_{||} - v_{||} = v'_{||},
\end{equation} because we can ignore the pulsar's overall line-of-sight motion.

We can then use some trigonometry to relate this quantity to the tangential velocity (noting that $\theta$ is very small): \begin{equation}
    \theta \approx \sin\theta = \frac{v'_{||}}{v_\perp} \approx \tan\theta = \frac{v_\perp \Delta t}{d},
\end{equation} which can be rearranged into the desired result:\begin{equation}
    a = \frac{\Delta v_{||}}{\Delta t} = \frac{v_\perp^2}{d}.
\end{equation}

\section{Derivation Two}

From Newtonian kinematics, we know that \begin{equation}
    d = \frac{1}{2}at^2 \vspace{0.1cm}
\end{equation} when the object has zero initial velocity. By converting $t$ into a small but finite interval of time $\Delta t$, this becomes \begin{equation}
    a = 2 \frac{\Delta d}{\Delta t^2}. 
\end{equation}

From Figure \ref{fig:schematic}, we see that \begin{align}
    (d + \Delta d)^2 &= (v_\perp \Delta t)^2 + d^2, \\
    2d \Delta d + \Delta d^2 &= v_\perp^2 \Delta t^2.
\end{align} We can neglect and discard the term that is quadratic in $\Delta d$ to obtain \begin{equation}
    2\frac{\Delta d}{\Delta t^2} = \frac{v_\perp^2}{d},
\end{equation} which is equal to $a$. 

\bibliographystyle{aasjournal}
\bibliography{references.bib}

\begin{thebibliography}{}
\expandafter\ifx\csname natexlab\endcsname\relax\def\natexlab#1{#1}\fi
\providecommand{\url}[1]{\href{#1}{#1}}
\providecommand{\dodoi}[1]{doi:~\href{http://doi.org/#1}{\nolinkurl{#1}}}
\providecommand{\doeprint}[1]{\href{http://ascl.net/#1}{\nolinkurl{http://ascl.net/#1}}}
\providecommand{\doarXiv}[1]{\href{https://arxiv.org/abs/#1}{\nolinkurl{https://arxiv.org/abs/#1}}}

\bibitem[{{Liu} {et~al.}(2018){Liu}, {Bassa}, \& {Stappers}}]{Liu2018}
{Liu}, X.~J., {Bassa}, C.~G., \& {Stappers}, B.~W. 2018, \mnras, 478, 2359, \dodoi{10.1093/mnras/sty1202}

\bibitem[{{Shklovskii}(1970)}]{Shklovskii1970}
{Shklovskii}, I.~S. 1970, \sovast, 13, 562

\end{thebibliography}

\end{document}